\documentclass[11pt]{article}
\usepackage{amssymb}

\newcommand{\inv}{^{-1}}
\newcommand{\rO}{\mathrm{O}}
\newcommand{\U}{\mathrm{U}}
\newcommand{\SU}{\mathrm{SU}}
\newcommand{\SL}{\mathrm{SL}}
\newcommand{\Gl}{\mathrm{Gl}}
\newcommand{\dg}{^\dag}
\newcommand{\pd}{'^\dag}

\begin{document}

\title{Conformal Symmetry on the Instanton Moduli Space}
\author{Yu Tian \\{\it Institute of Theoretical Physics, Chinese Academy of Sciences}\\
{\it P. O. Box 2735, Beijing 100080}\\
{\tt ytian@itp.ac.cn}}

\maketitle

\begin{abstract}
The conformal symmetry on the instanton moduli space is discussed
using the ADHM construction, where a viewpoint of ``homogeneous
coordinates" for both the spacetime and the moduli space turns out
to be useful. It is shown that the conformal algebra closes only
up to global gauge transformations, which generalizes the earlier
discussion by Jackiw et al. An interesting 5-dimensional
interpretation of the $\SU(2)$ single-instanton is also mentioned.
\end{abstract}

\newpage

We begin our discussion with a brief review of some relevant
aspects. For investigating conformal properties on the
4-dimensional spacetime, as well as for the study of instanton
solutions in gauge field theory and the expression of ADHM
construction, it is convenient to introduce a quaternionic
notation for the (Euclidean) spacetime coordinates:
\begin{equation}
x\equiv x^{n}\sigma_{n},\quad\bar{x}\equiv
x^{n}\bar{\sigma}_{n}=x^\dag,
\end{equation}
where $\sigma_{n}\equiv(\mathrm{i}\vec{\tau},1_2)$ and $\tau^i$,
$i=1,2,3$ are the three Pauli matrices, and the conjugate matrices
$\bar{\sigma}_{n}\equiv(-\mathrm{i}\vec{\tau},1_2)=\sigma_{n}^{\dag}$.
The action of the whole conformal group on the quaternionic
coordinates can be written elegantly as
\begin{equation}\label{conformal}
x\rightarrow\tilde{x}=(A x+B)(C x+D)\inv, \quad \det\left(%
\begin{array}{cc}
  A & B \\
  C & D \\
\end{array}%
\right)=1,
\end{equation}
where $A$, $B$, $C$ and $D$ are quaternions viewed here as
$2\times 2$ matrices. This fractional linear form of conformal
transformation enables us to introduce a kind of ``homogeneous
quaternionic coordinates" whose relation to the ordinary
quaternionic coordinates is
\begin{equation}
X\equiv\left(%
\begin{array}{c}
  u \\
  v \\
\end{array}%
\right), \quad x=u v\inv.
\end{equation}
Obviously, now the quaternionic matrix
\begin{equation}\label{linear}
\mathcal{C}\equiv\left(%
\begin{array}{cc}
  A & B \\
  C & D \\
\end{array}%
\right)
\end{equation}
acts linearly on $X$, which forms $\SL(2,\mathbb{H})$, the
Euclidean version of twistor representation of the conformal
group.

As is well known, the ADHM construction \cite{ADHM,CWS,KMS} (for
the self-dual instantons, without loss of generality) is described
by the basic object $\Delta$, an $(N+2k)\times 2k$ matrix linear
in the spacetime coordinates:
\begin{equation}\label{Delta}
\Delta=a+b x\equiv a+b(x\otimes 1_k),
\end{equation}
where the complex-valued constant matrices (in the canonical form)
\begin{equation}\label{canonic}
a=\left(\begin{array}{c}
K \\ \underline{a}
\end{array}\right),
\quad b=\left(\begin{array}{c}
0_{N\times 2k} \\ 1_{2k}
\end{array}\right).
\end{equation}
The degrees of freedom in $a$ constitute an over-complete set of
collective coordinates on the instanton moduli space. Roughly
speaking, $\underline{a}$ contains the information about positions
of the instantons, while $K$ contains that about sizes and gauge
orientations of them. Transformation (\ref{conformal}) gives
\cite{DHKM}
\begin{equation}
\Delta(\tilde{x};a,b)=\Delta(x;a D+b B,a C+b A)(C x+D)\inv.
\end{equation}
It can be seen that the $(C x+D)\inv$ factor is not essential, and
if $\Delta(x;a,b)$ satisfies the ADHM constraints, so does
$\Delta(x;a D+b B,a C+b A)$. Thus, we have the following conformal
transformation of the ADHM matrices:
\begin{equation}\label{original}
a\rightarrow a'=a D+b B, \quad b\rightarrow b'=a C+b A.
\end{equation}
In fact, this transformation can be seen more directly from the
following form of $\Delta$:
\begin{equation}
\Delta=\mathcal{A}X, \quad
\mathcal{A}\equiv\left(\begin{array}{cc} b & a
\end{array}\right),
\end{equation}
where $X$ takes the standard form
\begin{equation}
X=\left(\begin{array}{c}
x \\ 1_{2}
\end{array}\right)
\equiv\left(\begin{array}{c}
x \\ 1_{2}
\end{array}\right)\otimes 1_k,
\end{equation}
and the action (\ref{linear}) on $X$ as the anti-action on
$\mathcal{A}$, which can be regarded as the ``homogeneous
collective coordinates" on the instanton moduli space.

To achieve the canonical form (\ref{canonic}) from
eq.(\ref{original}), one has to make a symmetry transformation of
the ADHM construction, which takes $b'$ back to the canonical
form:
\begin{equation}\label{eq:b}
\Lambda b'G=b,
\end{equation}
where $\Lambda\in\U(N+2k)$ and $G\in\Gl(k,\mathbb{C})$, while $a'$
becomes
\begin{equation}\label{eq:a}
\tilde{a}=\Lambda a'G.
\end{equation}
Thus, the next step is to solve $\Lambda$ and $G$. First we have
\begin{equation}
(\Lambda b'G)\dg (\Lambda b'G)=G\dg b\pd b'G=b\dg b=1_{2k}.
\end{equation}
The solution of this equation is obviously not unique, but one can
naturally take a special one:\footnote{The square-root operation
in eq.(\ref{G}) is well defined, guaranteed by that $b\pd b'$ is a
positive-semidefinite Hermitian matrix. However, the validity of
the inversion operation here relies on the positive definiteness
of $b\pd b'$. It can be seen that this is the case in general.}
\begin{equation}\label{G}
G=G\dg=(b\pd b')^{-1/2}.
\end{equation}
It can be seen that this $G$ is in $\Gl(k,\mathbb{C})$, since
\begin{equation}
b\pd b'=C\dg a\dg a
C+A\dg\underline{a}C+(A\dg\underline{a}C)\dg+A\dg A
\end{equation}
and the matrix $a$ satisfies the ADHM constraints. Furthermore, we
have
\begin{equation}
(\Lambda b'G)(\Lambda b'G)\dg=\Lambda b'(b\pd b')\inv
b\pd\Lambda\dg=b b\dg=\left(%
\begin{array}{cc}
  0_N & 0 \\
  0 & 1_{2k} \\
\end{array}%
\right).
\end{equation}
To deal with matrix equations, especially for non-square matrices,
matrix partition is a rather general tool. In fact, the above
equation can be solved by the following decomposition:
\begin{equation}
\Lambda\dg=\left(%
\begin{array}{cc}
  \Lambda_{0\ (N+2k)\times N} & \Lambda_{1\ (N+2k)\times 2k} \\
\end{array}%
\right),
\end{equation}
where $\Lambda_0$ and $\Lambda_1$ are the orthonormal zero-mode
and nonzero-mode matrices of $b\pd$, respectively.

Suppose that $P$ is the projection operator onto the vector space
of nonzero modes of $b\pd$. It is easy to see
\begin{equation}
\Lambda_1\Lambda_1\dg=P=b'(b\pd b')\inv b\pd.
\end{equation}
So we can take
\begin{equation}
\Lambda_1=b'(b\pd b')^{-1/2}.
\end{equation}
There exists the completeness relation
\begin{equation}\label{complete}
\Lambda_0\Lambda_0\dg+P=1_{N+2k}.
\end{equation}
Again introduce the matrix partition
\begin{equation}
\Lambda_0=\left(%
\begin{array}{c}
  \lambda_{0\ N\times N} \\
  \underline{\Lambda}_{0\ 2k\times N} \\
\end{array}%
\right)
\end{equation}
and recall from eq.(\ref{canonic}) the partition
\begin{equation}
b'=\left(%
\begin{array}{c}
  K C \\
  \underline{a}C+A \\
\end{array}%
\right)\equiv\left(%
\begin{array}{c}
  L \\
  \underline{b}' \\
\end{array}%
\right).
\end{equation}
Now the completeness relation (\ref{complete}) becomes
\begin{equation}
\left(%
\begin{array}{cc}
  \lambda_0\lambda_0\dg+L (b\pd b')\inv L\dg & \mbox{h. c.} \\
  \underline{\Lambda}_0\lambda_0\dg+\underline{b}'(b\pd b')\inv L\dg &
  \underline{\Lambda}_0\underline{\Lambda}_0\dg+\underline{b}'(b\pd b')\inv\underline{b}\pd \\
\end{array}%
\right)=1_{N+2k}.
\end{equation}
This matrix equation is easily solved by (a special solution)
\begin{eqnarray}
  \lambda_0=\lambda_0\dg &=& [1-L (b\pd b')\inv L\dg]^{1/2}, \label{singular} \\
  \underline{\Lambda}_0 &=& -\underline{b}'(b\pd b')\inv
  L\dg\lambda_0\inv.
\end{eqnarray}
Here eq.(\ref{singular}) is actually the ``singular gauge" when
solving the matrix equation for the zero-mode matrix of $\Delta$
\cite{KMS}, while all other solutions of eq.(\ref{complete}) are
related to eq.(\ref{singular}) by $\U(N)$ gauge transformations.

From the above discussion we obtain
\begin{equation}
\Lambda=\left(%
\begin{array}{cc}
  \lambda_0 & -\lambda_0\inv L (b\pd b')\inv\underline{b}\pd \\
  (b\pd b')^{-1/2}L\dg & (b\pd b')^{-1/2}\underline{b}\pd \\
\end{array}%
\right).
\end{equation}
It is straightforward to check that equation (\ref{eq:b}) is
satisfied for this $\Lambda$ and $G$ in eq.(\ref{G}). Recall from
eq.(\ref{canonic}) the matrix partition
\begin{equation}
a'=\left(%
\begin{array}{c}
  K D \\
  \underline{a}D+B \\
\end{array}%
\right)\equiv\left(%
\begin{array}{c}
  M \\
  \underline{a}' \\
\end{array}%
\right).
\end{equation}
So from eq.(\ref{eq:a}) we have
\begin{equation}\label{moduli trans}
\tilde{a}=\left(%
\begin{array}{c}
  \lambda_0 M-\lambda_0\inv L (b\pd b')\inv\underline{b}\pd\underline{a}' \\
  (b\pd b')^{-1/2}(L\dg M+\underline{b}\pd\underline{a}') \\
\end{array}%
\right)(b\pd b')^{-1/2}.
\end{equation}

For the identity transformation, i.e., $A=D=1_2$ and $B=C=0$, it
is easy to see
\begin{equation}
\lambda_0=1, \quad \tilde{a}=\left(%
\begin{array}{c}
  K \\
  \underline{a} \\
\end{array}%
\right)=a,
\end{equation}
as expected. As the simplest nontrivial example, let us consider
the inversion transformation $\tilde{x}=x\inv$, i.e., $A=D=0$ and
$B=C=1_2$. Now we have $a'=b$ and $b'=a$, so the transformation
(\ref{moduli trans}) can be simplified to
\begin{equation}\label{inv}
\tilde{a}=\left(%
\begin{array}{c}
  -(1-W W\dg)^{-1/2}W \\
  1_{2k} \\
\end{array}%
\right)G\underline{a}\dg G,
\end{equation}
where $G=(a\dg a)^{-1/2}$ and $W=K G$. The composition of two
successive inversions (\ref{inv}) is an identity transformation on
$a$ (for nonsingular points of the moduli space), but it needs a
little effort to verify this fact, where two identities are
useful: one is\footnote{Refer to eqs.(42,43) in \cite{TZ} for a
proof of this identity.}
\begin{equation}
\underline{a}(a\dg a)\inv K\dg [1-K (a\dg a)\inv K\dg]\inv K (a\dg
a)\inv\underline{a}\dg+\underline{a}(a\dg
a)\inv\underline{a}\dg=1;
\end{equation}
the other is
\begin{equation}
1-\tilde{K}(\tilde{a}\dg\tilde{a})\inv\tilde{K}\dg=1-K (a\dg
a)\inv K\dg
\end{equation}
with $\tilde{K}$ defined by partition (\ref{canonic}) of
eq.(\ref{inv}).

Generically, it can be shown that the conformal transformations
(\ref{moduli trans}) of $a$ cannot form an exact realization of
the conformal group. However, they do form a ``projective"
realization. Suppose that the transformation matrices
$(\Lambda,G)$ and $(\Lambda',G')$ correspond to two successive
conformal transformations $\mathcal{C}$ and $\mathcal{C}'$ on $X$,
respectively. We have for the successive transformations on the
moduli space
\begin{equation}\label{successive}
\left(\begin{array}{cc} b & \tilde{\tilde{a}}
\end{array}\right)=\Lambda\left(\begin{array}{cc} b & \tilde{a}
\end{array}\right)\mathcal{C}G=\Lambda\Lambda'\left(\begin{array}{cc} b & a
\end{array}\right)\mathcal{C}'\mathcal{C}G'G,
\end{equation}
where the commutativity of $G'$ and $\mathcal{C}$ has been used.
At the same time, we have for the conformal transformation
$\mathcal{C}''=\mathcal{C}'\mathcal{C}$
\begin{equation}\label{composition}
\left(\begin{array}{cc} b & \hat{a}
\end{array}\right)=\Lambda''\left(\begin{array}{cc} b & a
\end{array}\right)\mathcal{C}''G''.
\end{equation}
It can be seen that eq.(\ref{eq:b}) fixes $\Lambda$ and $G$ up to
the following transformation:
\begin{equation}\label{trans}
\Lambda\rightarrow\left(%
\begin{array}{cc}
  \mathcal{U} & 0 \\
  0 & 1_2\otimes u \\
\end{array}%
\right)\Lambda, \quad G\rightarrow G u\dg,
\end{equation}
where $\mathcal{U}\in\U(N)$ and $u\in\U(k)$. Thus, we conclude
from eqs.(\ref{successive},\ref{composition}) that
$(\Lambda\Lambda',G'G)$ and $(\Lambda'',G'')$ must be related by
transformation (\ref{trans}). Considering the action (\ref{eq:a})
of $\Lambda$ and $G$ on $a$, $u$ is the so-called auxiliary
transformation and actually has no effect on the instanton moduli
space; but the $\SU(N)$ part of $\mathcal{U}$ acts nontrivially on
the instanton moduli space as a global gauge transformation. The
latter fact is in accord with the early work of Jackiw and Rebbi
\cite{JR}, who find that conformal transformations of the
single-instanton configuration are in some sense generically
accompanied by gauge transformations. If we disregard the global
gauge orientation of the instanton configuration, transformations
of the instanton moduli space induced by eq.(\ref{moduli trans})
will form a realization of the conformal group.

In some simple cases, one can easily go beyond singular gauge.
Here we use the inversion of $\SU(2)$ single-instanton as an
example. For an $\SU(2)$ single-instanton with size $\rho$ and
position $c$, we have the ADHM matrix
\begin{equation}
a=\left(%
\begin{array}{c}
  1_2\otimes\rho \\
  -c \\
\end{array}%
\right).
\end{equation}
Considering the inversion transformation, it is easy to see that
\begin{equation}
G=(|c|^2+\rho^2)^{-1/2}, \quad \Lambda=G\left(%
\begin{array}{cc}
  c & \rho \\
  \rho & -c\dg \\
\end{array}%
\right)
\end{equation}
is a (non-singular-gauge) solution of eq.(\ref{eq:b}). So we
obtain
\begin{equation}
\tilde{a}=\frac{1}{|c|^2+\rho^2}\left(%
\begin{array}{c}
  1_2\otimes\rho \\
  -c\dg \\
\end{array}%
\right),
\end{equation}
which looks as if the inversion transformation is performed in a
5-dimensional Euclidean space, with $\rho$ the 5th dimension. This
interesting phenomenon is also related to the well-known fact that
the single-instanton solution has an $\rO(5)$ invariance group
\cite{JR}.

\section*{Acknowledgments}

I would like to thank Dr Y. Zhang for helpful discussions. This
work is partly supported by NSFC under Grants No. 10347148.

\end{document}